\documentclass[epj]{svjour}
\usepackage{graphicx}
\usepackage{lipsum}

\begin{document}

\title{Neglecting Polydispersity Degrades Propensity Measurements in Supercooled Liquids}

\author{Cordell J. Donofrio \and Eric R. Weeks\thanks{\email{erweeks@emory.edu}}}
\institute{Department of Physics, Emory University, Atlanta, GA 30322, USA}

\titlerunning{Neglecting Polydispersity Degrades Propensity Measurements in Supercooled Liquids}
\authorrunning{Donofrio \& Weeks}

\date{\today}

\abstract{
We conduct molecular dynamics simulations of a bidisperse Kob-Andersen (KA) glass former, modified to add in additional polydispersity.  The original KA system is known to exhibit dynamical heterogeneity.  Prior work defined propensity, the mean motion of a particle averaged over simulations reconstructing the initial positions of all particles but with randomized velocities.  The existence of propensity shows that structure and dynamics are connected.  In this paper, we study systems which mimic problems that would be encountered in measuring propensity in a colloidal glass former, where particles are polydisperse (they have slight size variations). We mimic polydispersity by altering the bidisperse KA system into a quartet consisting of particles both slightly larger and slightly smaller than the parent particles in the original bidisperse system.  We then introduce errors into the reconstruction of the initial positions that mimic mistakes one might make in a colloidal experiment.  The mistakes degrade the propensity measurement, in some cases nearly completely; one no longer has an isoconfigurational ensemble in any useful sense.  Our results show that a polydisperse sample is suitable for propensity measurements provided one avoids reconstruction mistakes.
}


\maketitle

\section{Introduction}
\label{intro}

Many investigations have been conducted in the past half century to understand materials that, upon cooling, do not simply transition from an amorphous fluid state to an ordered solid state. These materials instead go through a glass transition wherein they maintain a disordered arrangement of molecules like a liquid, but have macroscopic physical properties akin to solids with a more crystalline structure. In particular, the viscosity of these materials grows exponentially as the molecular dynamics slow due to a relatively minor change in temperature.  In these samples as the glass transition is approached, the molecular dynamics are spatially heterogeneous \cite{kob97,donati98,poole98,karmakar14,richert02}, which is known to be a prevalent characteristic of the glassy slowing down in these materials. Specifically, at any moment there will be regions within the material where the particles exhibit slower mobility, compared to an average particle in the system, while other regions have relatively faster dynamics \cite{donati99,berthier04}. The sizes of the low and high mobility regions grow as the temperature continues to cool \cite{donati98}.

It has long been theorized that the local structure of the material plays a significant role in determining which areas would tend towards slower kinetics as the system evolves \cite{kivelson94,shintani06,coslovich11,royall15,schoenholz16}.  In 2004 Widmer-Cooper {\it et al}.~\cite{widmercooper04} provided evidence that the structure of the system was linked to its dynamics by introducing the concept of propensity. The idea is to simulate a system of particles many times over, having the particles always start from the exact same initial spatial configuration. This would create an iso-configurational (IC) ensemble of simulations. The difference in each simulation is that the initial velocities of the particles are randomized, consistent with the expected distribution of velocities given the temperature of the system.  In this way, they observed the trajectory of each particle many times starting from the same spatial configuration, but with no other memory of the previous state. They then defined propensity as
\begin{equation}
    p_i=\langle \Delta r_i^2 \rangle_{IC}
    \label{propensity}
\end{equation}
where $\Delta r_i$ is defined as the distance particle $i$ traveled over a specific time, and the averaging is done for the same particle over the iso-configurational ensemble. In particular the specific time chosen is typically $\tau_\alpha$, the relaxation time scale for which the higher mobility particles have (relatively) large displacements.  Widmer-Cooper {\it et al.} found that some particles have lower propensity:  these particles tend to travel less distance than a system wide average. Likewise, other particles have a high propensity value, and are more likely to move a large distance -- to rearrange.  Their conclusion is that indeed some aspect of the dynamics must be linked to the spatial structure.

However, this method does not identify exactly what details of the structure matter, and so subsequent work looked for structural quantities correlated with propensity.  The early studies were done with bidisperse systems:  mixtures of two particle sizes, to prevent crystallization.  Not surprisingly, particles belonging to the smaller species in the bidisperse mixture dominated the high propensity population \cite{widmercooper05}.  Later, several promising candidates for structural signals (free volume, size composition of neighbors, and initial potential energy) showed no significant relationship with propensity \cite{widmercooper05,widmercooper06,matharoo06}. Other work found that links between dynamics and local structures may be system dependent in binary atomic mixtures \cite{hocky14}. In a study of supercooled water, specific tetrahedral structures were found to correlate with low propensity regions \cite{malaspina09}. A study of a Lennard-Jones system found a connection between propensity and number of neighbors of various types of particles \cite{razul11}.

A collection of research began to focus on the fact that propensity was spatially correlated and that regions of high and low propensity were present throughout the system \cite{widmercooper04,widmercooper07}. That is, areas existed with a significant portion of high propensity particles and therefore these regions were more likely to undergo re-arrangement at $\tau_\alpha$ timescales. These domains could be found by coarse-graining the propensity values in the system and good results could be obtained at low temperature with IC ensembles consisting of as little as 50-100 simulations \cite{widmercooper07}. Of course this does not mean that such a region will always relax, as in order to do so the particles in such a region would have to move in a coordinated manner \cite{appignanesi07}. Indeed it was found that the propensity calculated for motions over $\tau_\alpha$ timescales were correlated with the Debye-Waller factor, which is a measure of the variability in particle position during the $\beta$-relaxation time scale (a shorter time scale than $\tau_\alpha$) \cite{widmercooper06}. Thus, longer term rearrangements are signaled by shorter term, coordinated motion in high propensity areas.

Some studies examined the connection between high-propensity regions and the potential energy landscape.  Propensity measured over a time corresponding to the maximum non-gaussian parameter, (also a $\beta$-relaxation timescale), was found to correlate well with motions over localized, unstable saddles in the potential energy landscape \cite{coslovich06}.  Other work showed that one can use the potential energy function to identify low-frequency, quasilocalized normal modes of vibration, and that the locations of these vibrational modes match well with higher propensity regions \cite{widmercooper08}.  Finally, the evolution of propensity itself occurs on intermediate time scales as meta-basin transitions occur \cite{appignanesi09,appignanesi09a}. 

All of these prior investigations have been simulation based, due to the fact that in order to measure propensity an iso-configurational ensemble has to be created:  the initial positions of all particles need to be realized many times. There would be great challenges presented if an attempt was made to create such an ensemble experimentally. However, if those challenges were to be overcome, then the calculation of propensity would be no more difficult than the calculation of other dynamical measures. A good candidate for attempting such an experiment is colloids.  These systems are comprised of small ($\sim 1$~$\mu$m diameter) solid particles in a liquid, which undergo Brownian motion \cite{hunter12rpp}.  Rather than controlling temperature, one controls the volume fraction, the fraction of volume occupied by the solid particles \cite{pusey86}.  As this is increased toward the glass transition volume fraction, the slowing down of the dynamics has a great many similarities with the glass transition in systems of small molecules or polymers \cite{hunter12rpp}.  A wide variety of investigations have shown that these materials are model glass formers  \cite{pusey86,vanmegen91,rosenberg89,marshall90,bartsch92,vanmegen94,hartl01}.  In particular, colloidal samples exhibit dynamical heterogeneity near the glass transition \cite{marcus99,kegel00,weeks00,latka09,mazoyer09}, the key quantity being probed by propensity measurements.  One could imagine initializing a set of colloidal particles, letting the system evolve long enough for it to have irreversibly rearranged itself, and then use a system of laser tweezers to bring the particles back to its original configuration. The Brownian motion of the solvent would ensure that initial movement away from this set configuration would be random. This should achieve physically what has only ever been simulated.  This would likely require a quasi-two-dimensional experiment, which is fairly common with colloids \cite{marcus99,cicuta03,konig07,ebert08,chen10,zhang11,vivek17,illing17}.  In particular, experiments by the Ganapathy group have successfully used laser tweezers to manipulate quasi-2D colloidal glass systems \cite{gokhale14,himanagamanasa15}, although only to pin particles rather than move them to specific locations.

The one physical reality that cannot be avoided is the fact that experimental colloids, unlike their simulated counterparts, are polydisperse:  as physical particles, they do not all have the same size.  The polydispersity is typically quantified by the standard deviation of the particle size distribution divided by the mean value, and it is often about 5\% \cite{bosma02,poon12,kurita12}. In this paper, we seek to understand how the measurement of propensity would be affected by having such diversity in particle size.  We do this by simulating the well-characterized bidisperse Kob-Andersen glass-forming system \cite{kob95a,kob95b}, altered by introducing polydispersity.  The polydispersity does not dramatically change the system, although it does slightly enhance the propensity signal (the variability between the lowest and highest observed propensities).  We then demonstrate that to reconstruct the initial state, it is not sufficient to merely bring back a similar size particle to an initial position; it is crucial to bring back the same particle to the initial position.  Returning to the original motivating question behind propensity \cite{widmercooper04}, this demonstrates that the structure-dynamics link must include particle size as part of what is meant by local structure.  This is fairly obvious for glasses composed of mixtures of small molecules or atoms with distinct and identical sizes (such as metallic glasses \cite{inoue11,zhang14,zhang15,kim15}), and a bit more intriguing and nontrivial for mildly polydisperse colloidal glasses.

\section{Methods}

\subsection{Simulations}
\label{simulations}

\begin{figure}[t]
\centerline{
\includegraphics[width=6cm]{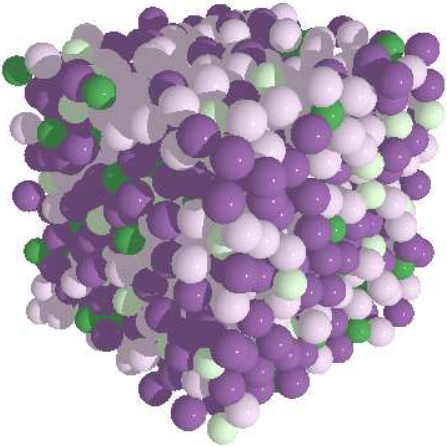}}
\caption{One of the ten initial configurations used to generate ensembles for $T=0.475$ and $\delta=1\%$. Purple shades are for species A particles while greens represent the smaller species B. Darker hues represent the plus variants of each species while lighter shades indicate the minus variants.}
\label{showsize}
\end{figure}

We start with the standard Kob-Andersen bidisperse glass former in 3 dimensions \cite{kob95a}. This mixture is governed by the Lennard-Jones (LJ) potential \cite{lennardjones31} which is of the form
\begin{equation}
\label{lj}
V_{LJ}=4\epsilon \bigg{[}\Big{(}\frac{\sigma}{r}\Big{)}^{12}-\Big{(}\frac{\sigma}{r}\Big{)}^{6}\bigg{]}.
\end{equation}
The interactions for both species in the system, denoted $A$ and $B$, are set by the parameters $\sigma_{AA}=1.0, \epsilon_{AA}=1.0, \sigma_{BB}=0.88, \epsilon_{BB}=0.5, \sigma_{AB}=0.8, $ and $\epsilon_{AB}=1.5$. Our mixture consists of $N_A$=800 and $N_B$=200 particles in a cube with periodic boundary conditions and sides of $9.4$, which matches the density used in Ref.~\cite{kob95a}. All distances are given in terms of $\sigma_{11}$, the time step is set to 0.005 for all simulations, time is given in reduced units of $(m\sigma_{11}^2/\epsilon_{11})^{1/2}$, and temperature is given in reduced units of $\epsilon\slash k_B$. Simulations are conducted with the LAMMPS \footnote{http://lammps.sandia.gov} software package, which uses the Verlet algorithm, and were done in the NVT regime~\cite{lammps1}. For temperatures of $T=1.0$ and $T=0.475$, we initialize 10 different particle configurations by running for $\tau=5\times 10^4$. We can be confident that all systems were equilibrated at this point as $\tau_\alpha \approx 5500$ is the structural relaxation time for our coldest system.


\begin{table*}[t]
    \centering
	\begin{tabular}{r@{\hspace{8pt}}l@{\hspace{20pt}}r@{\hspace{8pt}}l@{\hspace{20pt}}r@{\hspace{8pt}}l}
	$\sigma_{AA} = 1.00$ &  $\epsilon_{AA} = 1.00$ & $\sigma_{BB} = 0.88$ &  $\epsilon_{BB} = 0.50$ \hspace{20pt} & $\sigma_{AB} = 0.80$ &  $\epsilon_{AB} = 1.50$ \\ \hline

	$\sigma_{A_+A_+} = 1.01$ &  $\epsilon_{A_+A_+} = 1.01$ & $\sigma_{B_+B_+} = 0.8888$ &  $\epsilon_{B_+B_+} = 0.5050$ & $\sigma_{A_+B_+} = 0.8076$ &  $\epsilon_{A_+B_+} = 1.51425$  \\

	$\sigma_{A_-A_-} = 0.99$ &  $\epsilon_{A_-A_-} = 0.99$ & $\sigma_{B_-B_-} = 0.8712$ &  $\epsilon_{B_-B_-} = 0.4950$ & $\sigma_{A_-B_-} = 0.7924$ & $\epsilon_{A_-B_-} = 1.48575$  \\

	$\sigma_{A_+A_-} = 1.00$ &  $\epsilon_{A_+A_-} = 1.00$ & $\sigma_{B_+B_-} = 0.8800$ &  $\epsilon_{B_+B_-} = 0.5000$ & $\sigma_{A_+B_-} = 0.7724$ &  $\epsilon_{A_+B_-}  = 1.44825$  \\ \hline
	
	$\sigma_{A_+A_+} = 1.05$ &  $\epsilon_{A_+A_+} = 1.05$ & $\sigma_{B_+B_+} = 0.9240$ &  $\epsilon_{B_+B_+} = 0.5250$ & $\sigma_{A_+B_+} = 0.8380$ &  $\epsilon_{A_+B_+} = 1.57125$  \\

	$\sigma_{A_-A_-} = 0.95$ &  $\epsilon_{A_-A_-} = 0.95$ & $\sigma_{B_-B_-} = 0.8360$ &  $\epsilon_{B_-B_-} = 0.4750$ & $\sigma_{A_-B_-} = 0.7620$ & $\epsilon_{A_-B_-} = 1.42875$  \\

	$\sigma_{A_+A_-} = 1.00$ &  $\epsilon_{A_+A_-} = 1.00$ & $\sigma_{B_+B_-} = 0.8800$ &  $\epsilon_{B_+B_-} = 0.5000$ & $\sigma_{A_+B_-} = 0.6620$ &  $\epsilon_{A_+B_-}  = 1.24125$  \\
	
	 &  & & & $\sigma_{A_-B_+} = 0.9380$ &  $\epsilon_{A_-B_+} = 1.75875$ \\
	
	\end{tabular}
	\caption{Lennard Jones interaction parameters. The top row is for the Kob-Andersen binary \cite{kob95a}. The other sections are for the quartet system with $\delta=1\%$  (middle) and $\delta = 5\%$ (lower).  The left section of this table shows interactions between A type particles, the middle section is for B types, and the right section is for mixed AB interactions.}
	\label{ljparam}
\end{table*}

\subsection{Introduction of Polydispersity}

To study the effects of polydispersity, the standard Kob-Andersen bidisperse system must be altered so that there are more than just two particle sizes. While it would be ideal to draw particle sizes from a continuous distribution, if $n$ distinct sizes exist in our system we would need to define $n(n+1)/2$ LJ potentials. As a means of keeping computational times reasonable, we use a quartet system that contains a larger ($+$) and smaller ($-$) variant for each of the two original particle sizes. The magnitude of the variation is controlled by the parameter $\delta$. In having four different particle sizes, the interactions of the system are now governed by 10 distinct particle combinations and new LJ parameters of $\sigma$ and $\epsilon$ can be calculated using the following equations. For $X, Y\in A,B$ and $i,j \in +,-$ we define:

\begin{eqnarray}
\label{ssame}
\sigma_{X_+X_+} &=& (1+\delta)\sigma_{XX}, \\
\sigma_{X_+X_-} &=& \sigma_{XX}, \nonumber \\
\sigma_{X_-X_-} &=& (1-\delta)\sigma_{XX} \nonumber,
\end{eqnarray}
\begin{equation}
\label{sdiff}
\sigma_{A_iB_j} = 2(\sigma_{B_jB_j}+0.02\big)-\sigma_{Ai_iA_i},
\end{equation}
\begin{equation}
\label{epsilon}
\epsilon_{X_iY_j} = \epsilon_{XY}\frac{\sigma_{X_iY_j}}{\sigma_{XY}}.
\end{equation}


The form of eq.~\ref{sdiff} was chosen as it will produce the correct scaling used in the original bidisperse system for forces between $A$ and $B$ type particles. The rather favorable $AB$ interactions produced by these parameters discourages the original system from crystallizing.  Equation~\ref{epsilon} is the result of requiring the Lennard-Jones force, $F=\frac{dV}{dr}$, evaluated at the distance where the potential energy $V_{LJ}=0$, to be held constant within each column of Table ~\ref{ljparam}. This corresponds to changing the parent's $\epsilon$ value by the same percentage as the parent's $\sigma$ value was changed. A sample configuration at equilibrium is shown in fig.~\ref{showsize}. Here we see that the system is well mixed in that the different sizes are randomly distributed over the simulation volume.
Table ~\ref{ljparam} shows the values of $\sigma$ and $\epsilon$ for the binary system, a quartet with $\delta=1\%$, and a quartet with $\delta=5\%$.  The masses of all the particles are kept fixed at $m=1$, which is the same choice as the original Kob-Andersen system.  For comparison with colloids, this is a reasonable choice, as colloids are sufficiently small that their mass is not an important parameter for their dynamics, which are purely Brownian and thus mass-independent.

A useful consequence of eq.~\ref{ssame} is that the colloidal polydispersity of the $A$ and $B$ species in our mixture (defined as standard deviation of sizes $\sigma$ divided by mean size \cite{poon12}) is exactly equal to $\delta$. Of course, there is a qualitative difference between the bi-modal distribution that we are using for each species and a  continuous distribution, but numerically they are the same.

\begin{figure}
\includegraphics[width=8cm]{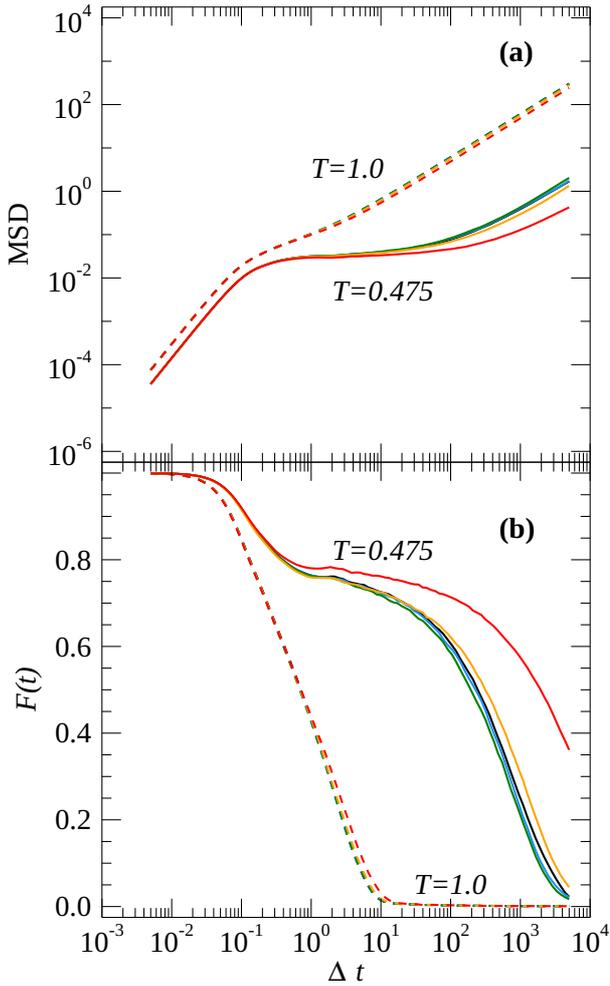}
\caption{\textbf{(a)} Mean squared displacement. Color indicates a $\delta$ value of $0\%$, $1\%$, $2\%$, $3\%$, and $5\%$ for dark blue, light blue, green, orange, and red respectively.  \textbf{(b)} Self part of the intermediate scattering function with same coloring scheme as panel (a). The wavevector used to calculate each $F(\Delta t;T,\delta)$ curve is taken by finding the $k$ value where the maximum $S(k)$ is observed for the corresponding $T$ and $\delta$ [see fig.~\ref{statst}(b) inset]. Values for $k$ range from 7.1 to 7.3.  For both (a) and (b), all curves represent the geometric average value of their respective function over the 10 different initial configurations we use.}
\label{dynst}
\end{figure}


Before we consider propensity measurements, we first wish to confirm that our quartet system is qualitatively similar to the original Kob-Andersen bidisperse system in terms of structure and dynamics.  Conceptually, we mimic how analysis would be done for an experimental system where polydispersity was ignored:  we treat the A and B particles separately, but do not distinguish between $+$ and $-$ variations.  In an experiment with a nominally bidisperse system, the A and B particles are presumed distinguishable but not necessarily the differences between particles of a given type \cite{cates15}.  The results below focus on the A particles (both $A_{+}$ and $A_{-}$) as is often done for this system. We initialize 10 separate systems for each value of $\delta$ and bring them to equilibrium as described above. Each system is then run for $\tau = 10^4$ in order to determine structural and dynamical quantities. 

We start with dynamics; in particular, we calculate the mean squared displacement and the self part of the intermediate scattering function for each system. Figure~\ref{dynst} shows the results for all $A$ type particles.  For both of these functions, we calculate the geometric mean for the set of 10 systems representing each $T$ and $\delta$ combination and it is these means which are displayed in fig.~\ref{dynst}. The geometric mean is used because for $T=0.475$ with $\delta=5\%$, there is a high level of variation seen in both functions produced by the individual configurations. By determining when $F(\Delta t; T,\delta)$ decays to $e^{-1}$ we define the relaxation time $\tau_\alpha(T,\delta)$. In our subsequent analysis, when treating a system with a given $T$ and $\delta$, we always use that system's specific $\tau_\alpha$. Again, the motivation here is to mimic what would be done in an colloidal experiment where one would measure $\tau_\alpha$ of the actual system rather than considering a hypothetical system where each member of a particle species has the exact same size. Additionally, we can now compare our new quartet system to the original binary. For both panels in fig.~\ref{dynst}, all data lie nearly on top of the the functions plotted from the binary case, which is shown in dark blue. The lone exception is when $\delta = 5\%$ and $T=0.475$. Under these conditions we see significant deviations indicating that the dynamics of the system have slowed considerably. 

\begin{figure}
\includegraphics[width=8cm]{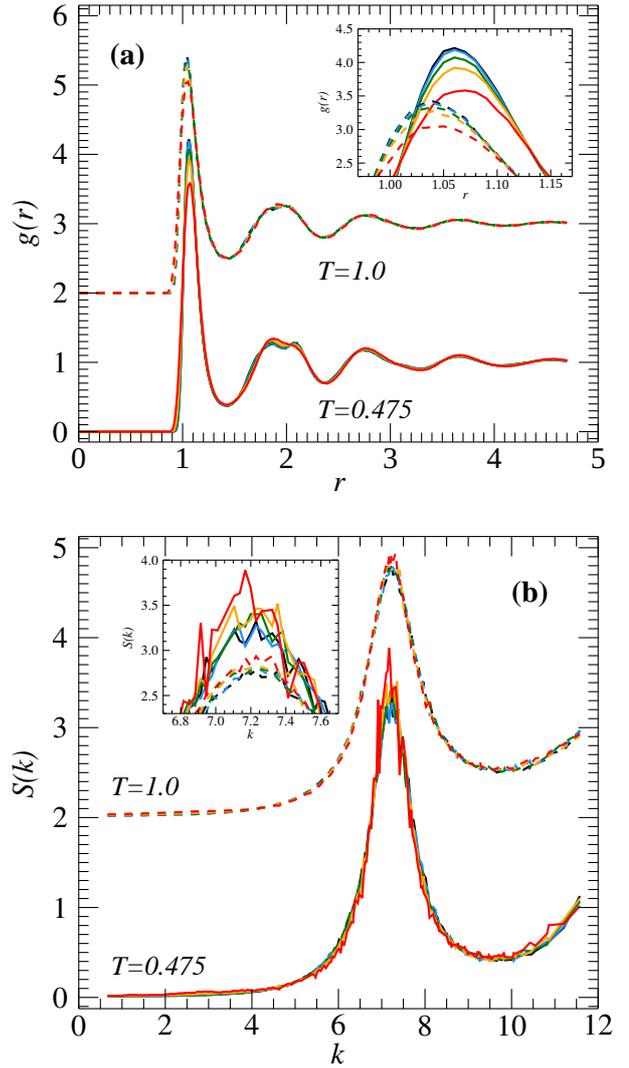}
\caption{\textbf{(a)} Radial distribution function for species $A$ particles; both $+$ and $-$ variants where applicable. Color indicates a $\delta$ value of $0\%$, $1\%$, $2\%$, $3\%$, and $5\%$ for dark blue, light blue, green, orange, and red respectively.\textbf{(b)} Static structure factor for all variants of species A particles. Coloring scheme is the same as panel (a). In both of the main panels the functions for $T=1.0$ have had 2 added to them so that they are separated from the lower temperature plots. The insets show the the first peak in $g(r)$ and $S(k)$ respectively. For the insets, the addition of 2 has been removed.}
\label{statst}
\end{figure}

To characterize static structure, fig.~\ref{statst}(a) shows the pairwise correlation function and fig.~\ref{statst}(b) shows the static structure factor; the data are for $A$ type particles and at all values of $\delta$ examined. Again, no distinction is made between the two variants of the $A$ particles in order to mimic calculations done in colloidal experiments. Similar to the dynamical measures of structure, these static quantities show little variation from the binary system except for when temperature is low, and $\delta$ is at $5\%$.

One possible explanation for why we see changes in the static and dynamical functions at high levels of $\delta$ is that as polydispersity increases, the difference between $A$ and $B$ type particles starts to become blurred. Notice that for $\delta=1\%$ in Table~\ref{ljparam}, all of the interaction distances $\sigma$ are greater for all possible $AA$ interactions than of any of the possible $BB$ interactions. These are in turn are greater than any of the $\sigma$ values for mixed species ($AB$) interactions, which implies that $A$ and $B$ particles prefer to be neighbors -- mixing is favorable. At higher values of $\delta$ the relative sizes become changed.  At $\delta=3\%$, $\sigma_{A_-B_+} > \sigma_{B_-B_-}$, (0.8828 as compared to 0.8536). At $\delta=5\%$, we have $\sigma_{A_-B_-} = 0.7620$ and $\sigma_{A_+B_-} = 0.6620$, meaning that these pairs of particle types have an even stronger preference to be neighbors, and potentially create regions of slower dynamics -- the cost of separating these pairs, in terms of requiring additional volume, are more severe.
We also have $\sigma_{A+A+}=1.050$, $\sigma_{A-A-} = 0.950$, and $\sigma_{B+B+}=0.924$, so the distinction between the $A+$ and $A-$ particles are stronger than the distinction between $A-$ and $B+$ particles.  In a colloidal setting this would represent the inability to distinguish a smaller variant of a `large' particle from a large variant of a `small' particle.

On the basis of figs.~\ref{dynst} and \ref{statst}, we conclude that the quartet system ($\delta > 0$) is reasonably similar to the original bidisperse system, with the possible exception of $\delta = 5\%$ which has slower dynamics.  All of these systems are reasonable glass-formers.

\section{Propensity Results}
\subsection{Simulation of Polydispersity}

\begin{figure}[b]
\includegraphics[width=8cm]{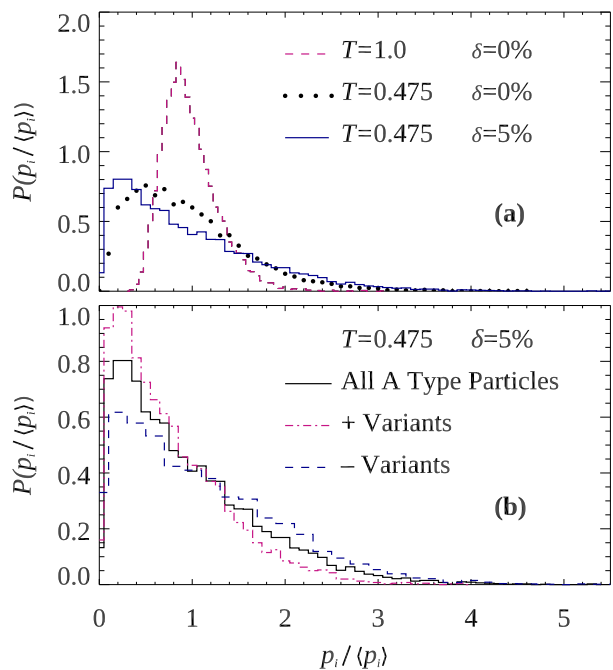}
\caption{Probability distributions of propensities for type A particles at the structural relaxation times for various systems as labeled. For all series, the data from each of the 10 ensembles is combined and normalized. \textbf{(a)} Normalization is done by dividing by the mean across all 10 ensembles that make up each individual distribution. The coefficient of variation for these distributions is 0.30 for dashed, 0.61 for dots, and 0.76 for solid. \textbf{(b)} The solid line is a reproduction of the solid line distribution in panel (a). The other series represent a breakdown of that low temperature/high polydispersity distribution into the two variant sizes ($A_+$ and $A_-$). Here each distribution is normalized by dividing by the mean value of the distribution that contains all size variants. The coefficient of variation for the dash-dot line $0.59$ and for the dashed line is $0.86$.
}
\label{pdist}
\end{figure}

After all systems are equilibrated, we construct an iso-configurational ensemble of 100 runs for each system. From this we are able to calculate the propensity $p_i$ of each particle (eq.~\ref{propensity}). While the value of propensity for any given particle indicates its own ability to move independent of the initial kinetics of the system, we are more interested in the distribution of propensities across the entire system. Figure~\ref{pdist}a displays the probability distribution of propensities for $A$ type particles under several conditions. Each condition is modeled using 10 separate ensembles and the propensity values for all ensembles are combined into a single distribution. In order to make comparisons between different conditions easier, the propensities are shown normalized by the mean value of the distribution. If the distribution is narrow, with all particles having nearly identical values, then measuring propensity is not conveying much information about the initial structures in the system. However, if there is a broad distribution of propensities, then it should be easier to find structures that either impede or facilitate mobility. The simulations at $T=0.475$ show that a wider variety of outcomes are seen compared to the higher temperature. At this low temperature, it becomes less probable to observe particles with propensities near the mean value of the system and more probable to find ones with both extremely low and extremely high propensities; propensity conveys more information.

\begin{figure}[t]
\includegraphics[width=8cm]{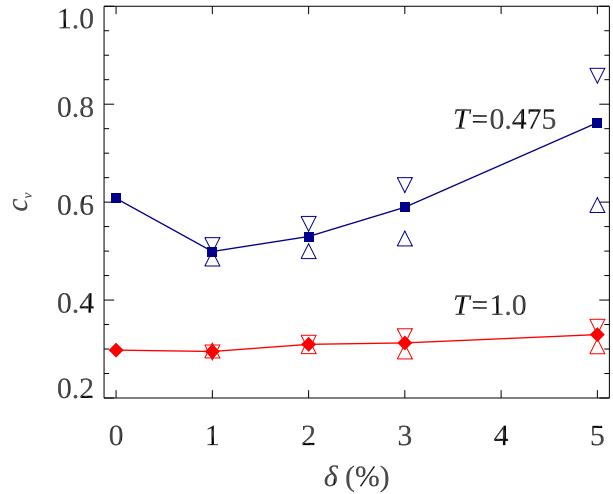}
\caption{Each filled symbol represents the average value of coefficient of variation in propensity of type $A$ particles at the structural relaxation time across the 10 different initial configurations at temperatures as indicated. The downward pointing open triangles correspond to the $c_v$ calculated just for the $A_-$ particles, and the upward pointing open triangles correspond to the $A_+$ particles.  In these cases $c_v$ was calculated by computing the standard deviation of the propensity of the individual particle type, and then dividing by the mean propensity for all $A$ particles.  Separate $\tau_\alpha$ are calculated for each $T$ and $\delta$ combination.} 
\label{spdvd}
\end{figure}

In order to quantify the width of any given distribution and thereby the heterogeneity of observed propensities, we calculate the coefficient of variation, $c_v$, of the propensity probability distributions. $c_v$ is defined as the standard deviation of a given distribution divided by the mean value of the same distribution; $\sigma_{p_i}/\langle p_i\rangle$.  A larger $c_v$ means that propensity is more ``interesting'' -- there is more difference between the low and high propensity particles. When calculating $c_v$ we only consider the $A$ type particles, though both the plus and minus variants are included.  The solid symbols in fig.~\ref{spdvd} show the coefficient of variation for various values of polydispersity at both $T=1.0$ and $T=0.475$. For the higher temperature, the introduction of polydispersity does not significantly change the propensity distribution width. This is unsurprising given the measurements in shown in fig.~\ref{dynst} for this temperature. The MSD curve does not display a glassy plateau suggesting that the dynamics are relatively spatially homogeneous across the system. We can conclude from this that the local structure does not play a significant role in determining the dynamics, and so modifying that structure by adding variation to the particle sizes should not have much effect on the dynamics or the heterogeneity of those dynamics.

However, at the lower temperature fig.~\ref{spdvd} shows that there is generally an increase in the coefficient of variation of the propensity as the polydispersity is increased. The bidisperse system is more glassy at this temperature, as evidenced by the plateau in the MSD curve, and so we would expect that altering the local structures at this temperature would affect the spatial heterogeneity of the dynamics. It appears that with the higher values of $\delta$ the system has become more glassy in that the MSD plateau is longer, and more extreme propensity values are observed. The solid line in fig.~\ref{pdist}a shows the distribution for $T=0.475$ and $\delta=5\%$, which has a notably long tail on the right with some particles displaying over five times the mean propensity value.

In general we find that these high propensity tails are dominated by the $A-$ variants, and that as $\delta$ increases, these smaller variants make up a larger percentage of all high propensity particles. Figure~\ref{pdist}(b) shows a breakdown of the propensity distribution for $T=0.475$ and $\delta=5\%$ by variant size, and the open symbols of fig.~\ref{spdvd} show the $c_v$ for the variant distributions. It is important to note that for these variant distributions, the normalization and the calculation of $c_v$ were done by dividing by the average value of all $A$ type particles, which makes for a more useful comparison. Figure~\ref{pdist}(b) indicates that the high propensity tail of the distribution consisting of all $A$ type particles is dominated by the $-$ variants. Correspondingly, fig.~\ref{spdvd} shows that the $c_v$ value for distributions containing only $-$ variants, (downward triangles), is higher at all $\delta$ values than the values for the $+$ variants, (upward triangles).

We are unable to explain the drop in the coefficient of variation observed from $\delta=0\%$ to $\delta=1\%$ for the $T=0.475$ data. To check the results we created a second set of 10 ensembles for $\delta=0\%$ for which we get a statistically similar value to the one plotted in fig.~\ref{spdvd}. Additionally, we created 10 ensembles where $\delta=0.001\%$. For this set of ensembles we found that $c_v=0.56$ which is also below the presented value for $\delta=0\%$. Lastly, the general trend observed is not sensitive to the time interval over which propensity was calculated. A similar trend existed at all late time scales including time of the maximum non-gaussian parameter for displacements $(t^*)$ \cite{kob97}. We do note that the same sequence can be observed in the self part of the intermediate scattering functions (ISF) shown in fig.~\ref{dynst}b. At late timescales, there is a slight drop in the ISF from $\delta=0\%$ to $1\%$ and $2\%$, but then a rise for $3\%$ and $5\%$. This correlation suggests a relationship between the two measurements (ISF and $c_v$).

\subsection{Simulation of Human Error}

One of the larger challenges that would have to be overcome if a propensity measurement were to be made on a colloidal system is the fact that this would involve resetting the physical system to its initial configuration many times, for example by using laser tweezers. We envision this to be a difficult process in part because of the polydispersity of the particles. We want to determine how measurements of propensity would be affected if the system is reset without regard to the specific size of each particle. That is, if an $A$ type particle exists at a specific location in the initial configuration, how important is it that the exact same particle is placed back at that location compared to an $A$ type with a slightly different size?

To investigate this, we alter the procedure that creates the iso-configurational ensembles. For each member of the ensemble, not only are the velocities of each particle randomized, but additionally a random subset of size $S$ are selected to be swapped with a different variant of the same species. That is, $A_+ \leftrightarrow A_-$ and $B_+ \leftrightarrow B_-$. Swaps between species were not examined. In each case the swaps are distributed with a $4:1$ ratio between $A$ and $B$ type particles in keeping with the ratio of those particles in the system. As an example, if $S=5$, then 5 pairs of particles will have their size variation swapped with four pairs being the $A$ type particles and 1 pair being $B$ types; thus ten particles will be ``incorrect'' in the reconstruction. The pairs that are chosen are random across each of the 100 simulations in the ensemble, so that for $S=5$ each particle would on average only be chosen to be part of a swapped pair once in the ensemble, $(100\times 5\times2/1000 = 1$). Ensembles are created with $S = 15, 30, 60,125, 250$. When 250 swaps are made then exactly half of the particles will be involved in each swap, as $N=1000$, making this the condition under which we achieve the highest level of random changes to the system. The inclusion of this maximal level of randomization simulates an experimenter that attempts to construct an iso-configurational ensemble ignoring any difference in particle size due to polydispersity. Assembly in such a manner would be considerably easier, and we wish to understand whether such a shortcut could be justified.

\subsection{Results from Introduction of Error}
\begin{figure}[t]
\includegraphics[width=8cm]{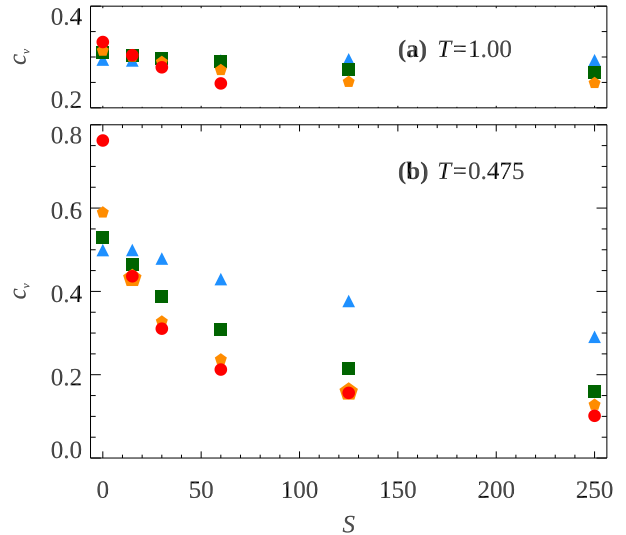}
\caption{Coefficient of variation of propensity values (for $A$ particles and at $\tau_\alpha$) as a function of the number of swaps. Color and shape indicates $\delta$ values of $1\%$, $2\%$, $3\%$, and $5\%$ for blue triangles, green squares, orange pentagons, and red circles respectively. }
\label{spdvs}
\end{figure}

Figure~\ref{spdvs} shows the effect that the number of swaps has on the coefficient of variation of propensity across the system. The top panel shows this for the higher temperature of $T=1.0$. Similar to the results of increasing polydispersity, increasing the number of swaps made at this temperature only causes a slight decrease in the coefficient of variation. Thus the values of propensity observed in these systems appears to be little changed by the introduction of swaps, even for larger amounts of polydispersity (larger $\delta$).  Of course, the low $c_v$ for the $\delta=0$ system indicates that at this temperature the dynamics of the system are fairly homogeneous. There is little information to be found in the local structures of the system to give insight into the dynamics. Thus there is little information to be lost even when many structural changes occur due to the swapping of highly polydisperse particles.

Data are unable to be collected for $S$ values of 125 and 250 for the highest polydispersity level of $\delta=5\%$. Given the higher temperature, particles can be found farther from the Lennard-Jones potential minimum while the system is at equilibrium, and in particular some particle pairs are closer together than the distance that minimizes their potential energy. When a large number of swaps occur, it becomes very probable that a $+$ variant is moved to a position where a $-$ variant had strayed far from the potential minimum and close to a $+$ variant. The resulting force on this $++$ particle pair jumps several orders of magnitude, resulting in extraordinary velocities and the eventual failure of the software to be able to keep track of at least one of the particles. As it became probable that multiple simulations in the ensemble would lose particles, we only present data for the lowest four values of $S$ for $\delta=5\%$. Data for the missing points could possibly be collected if the time step of the simulations is made smaller, but seems unlikely to affect the conclusions.  We assume that in a colloidal system, mistakes causing such an increase of force would be readily apparent and could be corrected.

\begin{figure}[t]
\includegraphics[width=8cm]{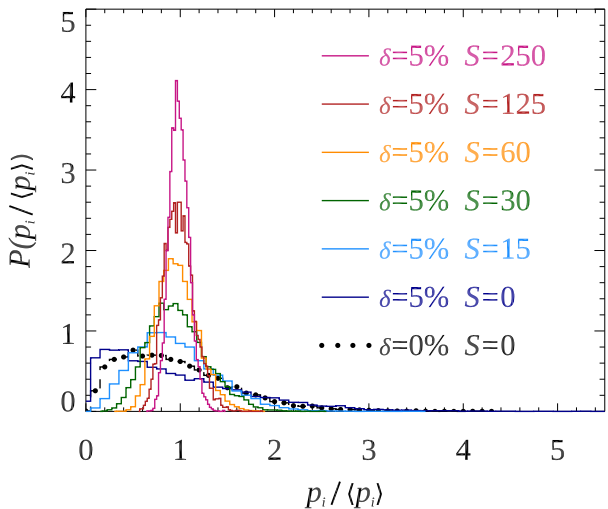}
\caption{Probability distributions of propensity of type $A$ particles and the effect of introducing errors into the construction of the iso-configurational ensembles. The distributions for no errors, $S=0$, with $\delta=0\%$ and $\delta=5\%$ are reproduced from fig.~\ref{pdist}a as the dotted line and the solid dark blue line respectively. As the number of errors increases, the distributions become narrower, indicating a decrease in the ability for the initial structure to clearly predict future dynamics.
}
\label{pdist1}
\end{figure}

\begin{figure}[b]
\centerline{\includegraphics[width=8cm]{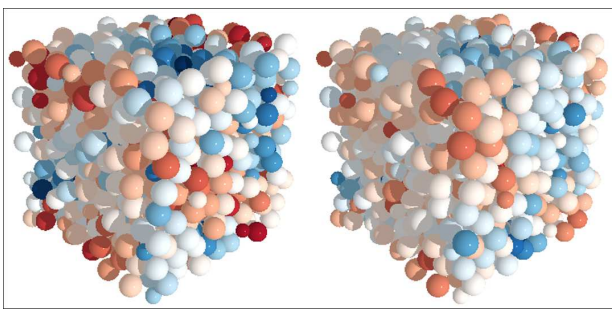}}
\caption{The color of each particle indicates the particle's propensity as compared to the system mean propensity, for a system with zero swaps (left) and 250 swaps (right). The redder a particle appears the higher its level of propensity, while blue indicates lower propensity particles. White particles experience the mean level of propensity. Both systems have $T=0.475$, $\delta=1\%$, and have the exact same particle configuration, highlighting how accidental mistakes (the swaps) decrease the propensity signal present in the zero-mistakes system (left).}
\label{3dpair}
\end{figure}

\begin{figure*}
\includegraphics[width=17cm]{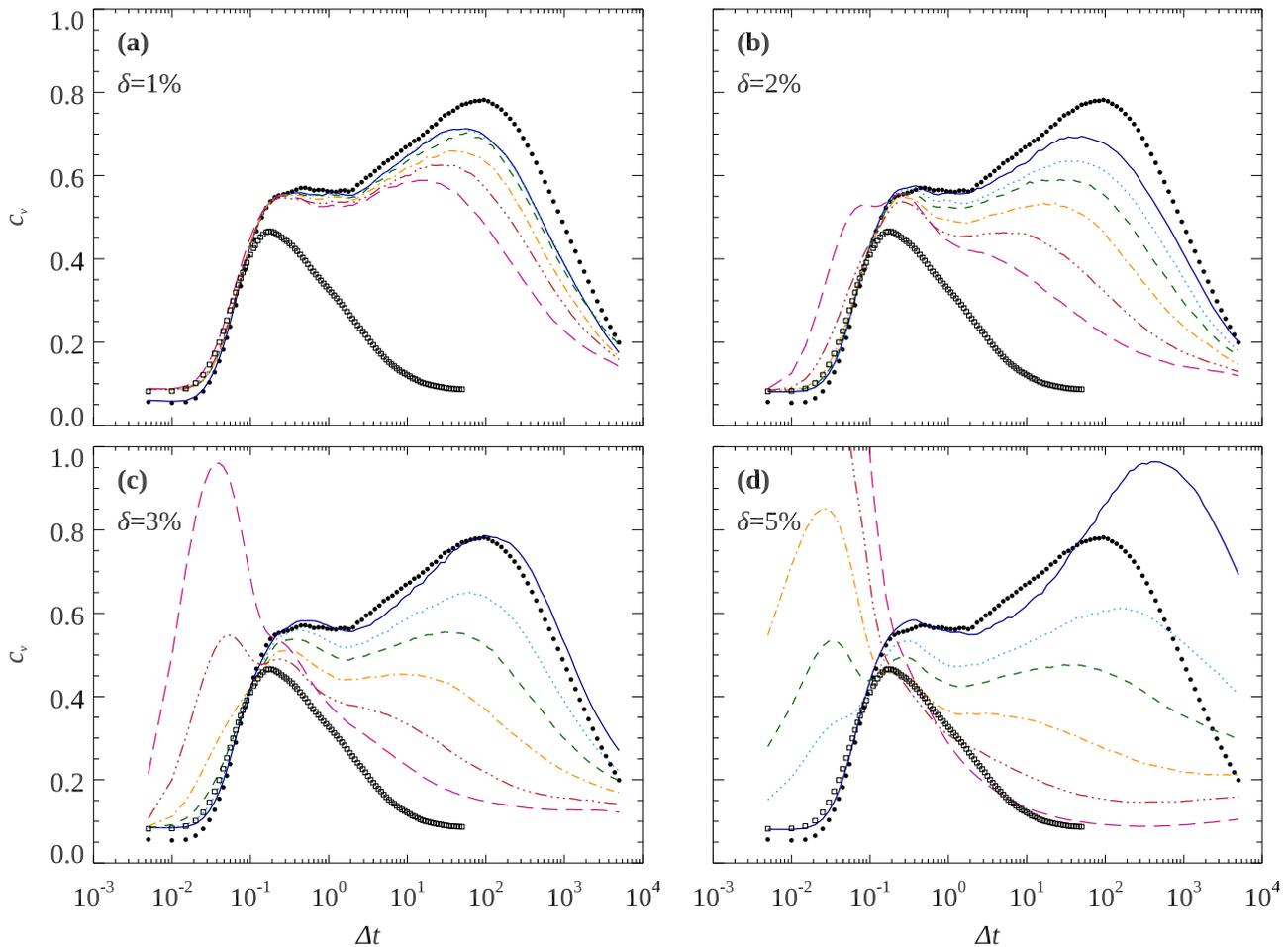}
\caption{The coefficient of variation for propensity as function of time. All panels represent simulations done at $T=0.475$, but for different levels of polydispersity in the quartet, $\delta$. The solid circles in each plot show the function for the original KA binary mixture. The dark blue solid line in each panel shows the $S=0$ data for simulations in which no swaps were done when constructing the iso-configurational ensemble. In (a), the number of swaps $S$ increases as 15, 30, 60, 125, and 250 as the curves decrease below the dark blue solid line ($S=0$); the colors and line styles are identical in the other panels. The open circles are for a binary mixture at $T=1.0$.}
\label{cvvt}
\end{figure*}

The lower panel of fig.~\ref{spdvs} shows results for a temperature of $T=0.475$. Here we see that as the number of swaps is increased, the coefficient of variation of the propensity decreases dramatically. This is true for all polydispersity levels, but there is a greater effect for larger values of $\delta$. Figure~\ref{pdist1} shows the probability distribution of propensities for $T=0.475$, $\delta=5\%$, and all amounts of swaps tested. For comparison, the distribution for the same temperature in the original Kob-Andersen binary ($\delta=0$) is reproduced here. As the number of mistakes increases, with this level of polydispersity, the distribution of propensities narrows:  both high and low propensity particles are lost. Thus the dynamics of the system are (apparently) becoming more uniform as we introduce mistakes in assembling the iso-configurational ensemble. Note that the presence of these mistakes could cause the relationship between temperature and the coefficient of variation to be viewed incorrectly. In fig.~\ref{spdvd} we see that at all levels of $\delta$, $c_v$ is larger for lower temperature systems:  propensity becomes more significant as $T$ approaches the glass transition $T_g$. However, $c_v = 0.24$ for ($T=1.0,\delta=5\%, S=60$), while $c_v=0.21$ for the same conditions at $T=0.475$. All points with $\delta \ge 3\%$ and $S \ge 60$ also show this inverted relationship between $c_v$ and $T$.

The uniformity of the dynamics caused by large numbers of swaps can also be seen in fig.~\ref{3dpair} where color represents $p_i/\langle p_i \rangle$. Both renderings are for systems at $T=0.475$, $\delta=1\%$, and the ensembles which produced both data sets stem from the same initial configuration. The system on the left has had no swaps made when determining propensity, while 250 swaps were made for the system on the right, thus maximum randomness when constructing the isoconfigurational ensemble. Keep in mind that the propensities are calculated from an ensemble of 100 simulations, and that in each of those simulations particles were chosen at random to be swapped. So for the image on the right, each particle had a 50/50 chance of being selected for a swap in each simulation run. The uniformity in color for the swapped system, in comparison to the non-swapped system, clearly shows that the measured propensity (right) appears to be more uniform throughout the system -- as compared to the true propensity (left).

Figure~\ref{cvvt} shows the coefficient of variation as a function of time for the simulations conducted at $T=0.475$, with each of the four panels representing a different level of polydispersity. This data was collected by calculating the propensity of the particles over 100 different time scales, while previously we have only discussed propensity for $\tau_\alpha$.   In each plot, the different lines represent simulations where different numbers of swaps occurred. Panel (a) shows the data for $\delta=1\%$. Notice that there is a small jump in the initial value of $c_v$ for all values of $S$, indicating that when even a small number of mistakes occur during reassembly, there is a wider variety of displacements happening in the first step of the simulation. We assume that this is due to the fact that any time swaps occur, the equilibrium of the system is disturbed and we are straying from the iso-configurational ensemble. It is possible that the inherent structure \cite{stillinger83} of the system has been changed, though we did not calculate this. However, at this low level of polydispersity, all of the curves follow a similar shape and so we conclude that the evolution of the material is similar in nature to a pure binary mixture. As more swaps are introduced, displacements become more uniform at longer timescales.

\begin{figure}[htbp]
\centerline{\includegraphics[width=6cm]{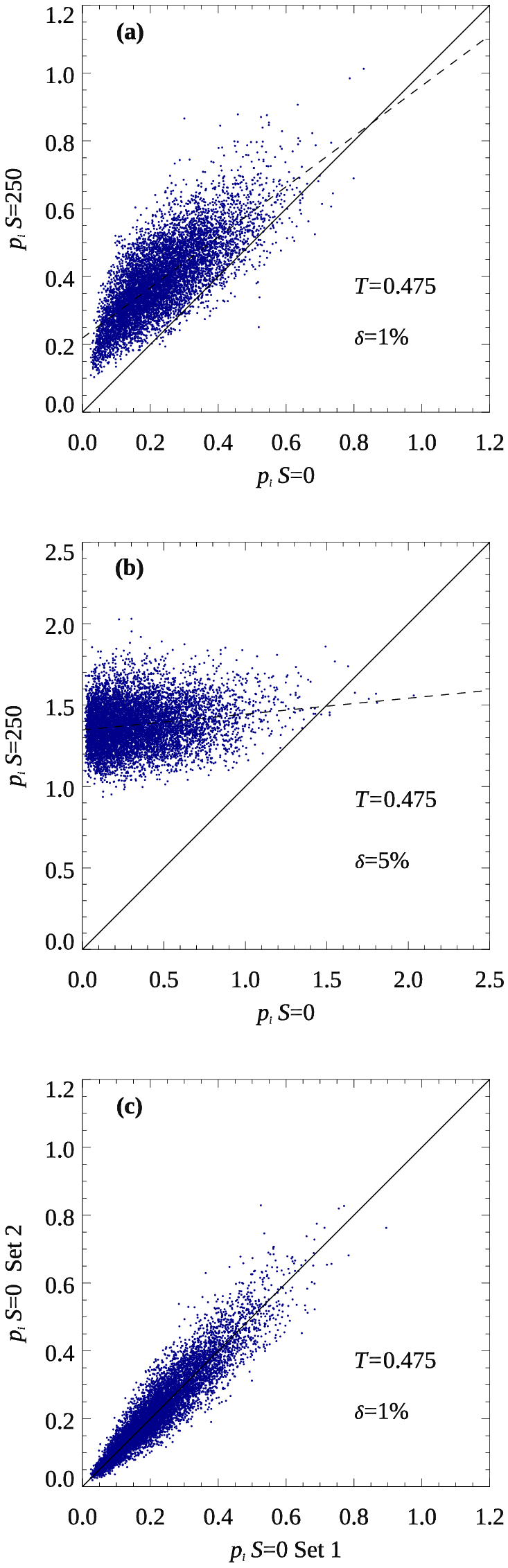}}
\caption{Panels (a) and (b) show plots which compare a particle's propensity when it is in an ensemble with 250 swaps (maximal random errors) to an ensemble with 0 swaps (the true propensity). The solid line indicates $x=y$, marking particles that experience no effect from the swaps. The dashed line indicates a line of best fit to the data. The levels of polydispersity are different with $\delta=1\%$ in (a) and $\delta=5\%$ in (b). The last panel, (c), also contains data with $\delta=1\%$ but here the true propensity from one ensemble is compared to the true propensity from a second ensemble of equal size, demonstrating the variability in propensity due to the finite number (100) of configurations generated for the isoconfigurational ensemble. $T=0.475$ in all three plots.}
\label{p250vp0}
\end{figure}

Of course as we introduce more size variation and simulate more errors in reconstructing the ensemble, the initial rise in $c_v$ becomes more pronounced. Under more extreme conditions such as $\delta \geq 3\%$ with $S \geq 125$ we see very large rises in $c_v$ at early time scales as the swapping causes some particles to be very near each other resulting in large force values.  It is these extreme outliers that raise the $c_v$.  Additionally, the general shape of the time evolution has changed completely as the secondary rise in $c_v$ is suppressed or even non-existent. We conclude that in these cases the swapping is enough to render the idea of an iso-configurational mostly meaningless as displacements across the ensemble become more uniform.

We examine this further in fig.~\ref{p250vp0} by comparing the propensity value for each particle in systems where no swaps occur to the propensity of the same particle in ones where many swaps occur. The data in panel (a) are for simulations at $T=0.475$ with $\delta=1\%$ just as in fig.~\ref{3dpair}, except here we include data from all 10 initial configurations and do not normalize by the mean. The horizontal axis is the propensity for a given particle when no swaps are made -- the true propensity that we desire to measure. The vertical axis is the propensity for the particle that exists at the same location in the iso-configurational ensemble, but for the situation using $250$ swaps; in other words, when the maximum number of random mistakes is realized. The plot shows that there is a relatively high level of correlation between the two data sets:  particles that have higher levels of propensity in the original system tend to still have a high level of propensity in the swapped system -- keeping in mind that in the swapped system, a ``particle'' with a certain propensity now corresponds to the {\it position} of a particle prior to the swapping, as the literal particle may or may not stay in that location. This indicates that the structural information is still present despite the large number of swaps that were made, though variation in the plot indicates that this information may be harder to discover. As a control, fig.~\ref{p250vp0}(c) is a similar plot, but here the true propensity derived from one ensemble is plotted against the true propensity from a second ensemble where both sets had the same temperature and level of polydispersity. The correlation between the two sets is 0.93 and this gives a sense of the amount of spread present in the data due to the inherent fluctuations from constructing iso-configurational ensembles with randomized initial velocities.  If one were to construct more than 100 realizations in the iso-configurational ensemble, the variability in these data would converge toward the $x=y$ line.

The relationship between the true propensity and propensity measured with maximum swaps is much more muddled in fig.~\ref{p250vp0}(b), which is the same type of plot but for a system with $\delta=5\%$. Here there is very little correlation between the propensities in the original and the swapped systems, indicating that any structural information conveyed by the propensity in the original system has been lost. It seems that while some particles have higher or lower propensities in the system with swaps, that has no correlation with having higher or lower true propensity when there are no swaps. The measured propensity is apparently due to the swapping algorithm itself, rather than the original structure. For example, as discussed above, if a swapped particle moves into a position where it experiences a dramatically larger force immediately after the swap, that might result in a larger displacement and thus (every time it is swapped) a larger measured propensity.  But the presence of this sort of swap-error-induced propensity tells us nothing about the intrinsic dynamics of the original system; this is not the structure-dynamics relationship we are looking for.

\begin{figure}[htb]
\includegraphics[width=8cm]{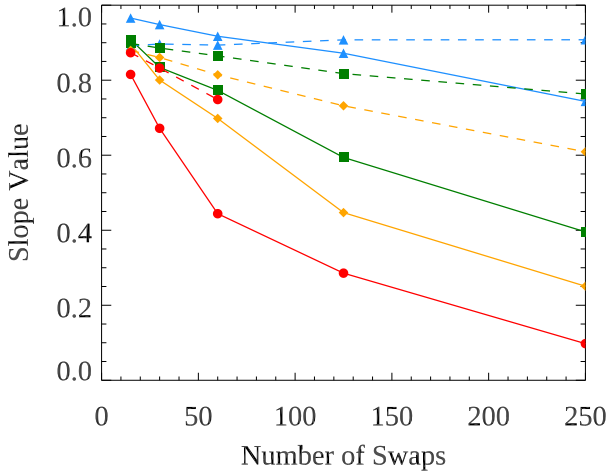}
\caption{The slope values from when propensity with a given number of swaps was plotted against the propensity with zero swaps. Each line is a particular $T$ and $\delta$ combination. All solid lines are $T=0.475$ and all dashed lines are $T=1.0$. $\delta=$ 1\%, 2\%, 3\%, and 5\% are respectively dark blue, light blue, green, and red.}
\label{swapslopes}
\end{figure}

We wish to quantify how polydispersity $\delta$, swapping errors $S$, and temperature $T$ interplay. To do this, lines of best fit are found for the data in panels (a) and (b) of fig.~\ref{p250vp0}, which are shown as dashed lines, as well as for other values of $\delta$, $S$, and $T$. A slope of 1, depicted in the plots as solid lines, would indicate good predictability between the true propensity and the measured propensity. However, these plots show that the slope is less than 1, indicating less correlation between the measured and true propensity values. The summary of all our data is shown in fig.~\ref{swapslopes}, which has the slopes for all systems. Again, we see that at $T=1.0$ (dashed lines), not much structural information is lost even for large numbers of swaps. For $\delta=3\%, S=250,$ and $T=1.0$, the lowest slope value found is $\approx 0.6$. For the lower temperature data (solid lines), we see the drop off can be quite severe, with the worst case (corresponding to panel (b) of fig.~\ref{p250vp0}) having a slope of $\approx 0.1$.  At this point nearly all of the apparent propensity is fictitious, with little correlation to the propensity one wishes to measure.

While a drop in slope value indicates that structural information is scrambled by the introduction of errors, it is also clear that these mistakes cause the measured propensity values to increase in general. In both fig.~\ref{p250vp0}(a) and (b), the data lie above the $y=x$ line. Compare this with the plot in fig.~\ref{p250vp0}(c) where the data appear equally above and below the line. This general increase in the measured propensity values appears to correlate with the level of polydispersity, as the increase is clear but mild in fig.~\ref{p250vp0}(a) where $\delta=1\%$ but is quite pronounced in fig.~\ref{p250vp0}(b) where $\delta=5\%$. If a high value of propensity is supposed to indicate a region in the material that is likely to be the site of a re-arrangement in the near future, then an experiment conducted with high levels of polydispersity, at low temperature, and with an assumption of indistinguishably between members of each particle species, then the results would be incorrectly interpreted as a majority of the system being prone to reorganization.  We reiterate that swapping generally causes all particles to have moderate to high values of propensity.  This is the reason why the slope of the best fit line decreases, in other words, why the slopes plotted in Fig.~\ref{swapslopes} are all below 1, because the true low propensity particles have their propensities increased with swapping, while the true high propensity particles generally keep a high propensity value with swapping.

\section{Conclusions}

The most challenging part of attempting a propensity measurement with a colloidal system will inevitably be determining a way to produce an iso-configurational ensemble. The difficulties faced in re-assembly will only be made worse by the fact that real particles will exhibit polydispersity.  We have shown that the intrinsic polydispersity to colloids will not be an insurmountable obstacle as long as care is taken to minimize mistakes when reassembling the system. At a temperature of $T=0.475$ we see that even a mistake rate of $3\%$, ($S=30$), would result in the coefficient of variability of the propensity dropping from 0.76 to 0.31 for a system with $\delta=5\%$. Indeed, at a polydispersity level of $5\%$, the idea of an iso-configurational ensemble appears to be broken with even the occasional error in reconstruction.

There are two potential goals one could investigate in a colloidal experiment.  One may want to show that propensity becomes more significant as the glass transition is approached, and as we have argued above, a way to do this is to measure $c_v$:  the coefficient of variation, where large values indicate propensity has more ``signal.''  Another goal would be to look for structure-dynamic relations, in which case one needs to measure individual propensity values accurately.  For both of these goals, there are at least two potential solutions.  First, one could use colloidal particles of low polydispersity, although those are difficult to find \cite{poon12}.  Our data indicate $\delta < 3$\% is necessary.  Second, one can work hard to achieve $S=0$:  ensuring that every particle is returned to its specific initial position.  If you have a choice, this second option is the better choice.  Our results show that even a polydisperse system ($\delta=5$\%) has reasonable dynamics and a nicely measurable propensity, so long as one avoids reconstruction errors.

\begin{acknowledgement}
This work was supported by the National Science Foundation (Grant No. DMR-1609763).
\end{acknowledgement}

Author contributions:  C.D. and E.R.W. designed the project; C.D. conducted the simulations, analyzed the data, and prepared the figures; C.D. and E.R.W. wrote the paper.

\bibliography{cory}
\bibliographystyle{epj}
\end{document}